\documentclass[12pt]{article}
\usepackage[dvips]{graphicx}

\begin{document}

\title{Electric dipoles on magnetic monopoles\\in spin ice}

\author{D. I. Khomskii\\[\medskipamount]
\normalsize II. Physikalisches Institut, Universit\"at zu K\"oln,\\[-\smallskipamount]
\normalsize Z\"ulpicher Str. 77,\\[-\smallskipamount]
\normalsize 50937 K\"oln, Germany}

\date{}

\maketitle

\begin{abstract}
\noindent
The close connection of electricity and magnetism is one of the
cornerstones of modern physics. This connection plays crucial role
from the fundamental point of view and in practical
applications, including spintronics and
multiferroic materials.
A breakthrough was a recent proposal
that in magnetic materials called spin
ice the elementary excitations have a magnetic charge and behave as magnetic monopoles.
I show
that, besides magnetic charge, there should be an electric dipole attached
to each magnetic monopole.
This opens new possibilities to study and to control such monopoles by electric field.
Thus the electric--magnetic analogy goes even
further than usually assumed: whereas electrons
have electric charge and magnetic dipole (spin), magnetic
monopoles in spin ice, while having magnetic charge, also have
electric dipole.
\end{abstract}

\section*{Introduction}
Spin ice materials present a very interesting class of magnetic materials
\cite{bramwell}. Mostly these are the pyrochlores with strongly
anisotropic Ising-like rare earth such as Dy or Ho \cite
{revmodphys}, although they exist in other structures, and one cannot exclude that similar materials
could also be made on the basis of transition metal elements with
strong anisotropy, such as
Co$^{2+}$ or~Fe$^{2+}$. Spin ice systems consist of a network of corner-shared
metal tetrahedra with effective ferromagnetic coupling between spins~\cite{hertog, yavorskii},
in which in the ground state the Ising spins are
ordered in two-in/two-out fashion. Artificial spin ice systems with different structures
have also been made~\cite{wang, tchernysh, ladak, mengotti}.

Spin ice systems are {\it bona fide} examples of frustrated systems, and they
attract now considerable attention, both because they are interesting in their own right
and because they can model different other systems, including real
water ice \cite{pauling}. A new chapter in the study of spin ice was
opened by the suggestion that the natural
elementary excitations in spin ice materials --- objects with
\hbox{3-in/1-out} or \hbox{1-in/3-out} tetrahedra --- have a magnetic charge \cite{ryzhkin}
and display many properties similar
to those of magnetic monopoles~\cite{castelnovo}.
Especially the last
proposal gave rise to a flurry of activity, see e.g.\ \cite{gingras2009},
in which, in particular, the close analogy between electric
and magnetic phenomena
was invoked.
Thus, one can apply to their
description many notions developed for the description of systems of charges
such as electrolytes; this description proves to be very efficient
for understanding many properties of spin ice.

Until now the largest attention was paid to the magnetic properties
of spin ice, both static and dynamic, largely connected with monopole excitations
\cite{morris, fennel, kadowaki, jaubert, slobinsky, giblin}, and the main tool to modify their properties was
magnetic field, which couples directly to spins or to the magnetic
charge of monopoles. I argue below that the magnetic monopoles in
spin ice have yet another  characteristic which could allow for
other ways to influence and study them: each magnetic monopole, i.e.\
the tetrahedron with \hbox{3-in/1-out} or \hbox{1-in/3-out} configuration, shall
also have an electric dipole localized at such tetrahedron. This
demonstrates once again the intrinsic interplay between magnetic and
electric properties of matter.

It is well known that some magnetic textures can break inversion
symmetry -- a necessary condition for creating electric dipoles.
This lies at the heart of magnetically-driven ferroelectricity in
type-II multiferroics~\cite{trends}. There exists, in particular,
a purely electronic mechanism for creating electric dipoles. I
demonstrate that a similar breaking of inversion symmetry, occurring in
magnetic monopoles in spin ice, finally leads to the creation of
electric dipoles on them.

\section*{Results}
\subsection*{The appearance of dipoles on monopoles}
The usual description of magnetic materials with localized magnetic
moments is based on the picture of strongly correlated electrons with
the ground state being a Mott insulator, see e.g.\ Ch.~12 in
\cite{khomskiibook}. In the simplest cases, ignoring orbital effects
etc., one can describe this situation by the famous Hubbard model
\begin{equation}
{\cal H} = -t\sum_{\langle ij\rangle,\sigma}c^\dagger_{i\sigma}c^{\vphantom+}_{j\sigma} + U\sum_i n_{i\uparrow}n_{i\downarrow}\;,
\end{equation}
where $t$ is the matrix element of electron hopping between
neighbouring sites $\langle ij\rangle$ and $U$ is the on-cite Coulomb repulsion. For
one electron per site, $n = N_e/N = 1$, and strong interaction $U\gg t$ the
electrons are localized, and there appears an antiferromagnetic
nearest neighbour exchange interaction $J = 2t^2/U$ between localized
magnetic moments thus formed (which acts together with the usual
classical dipole-dipole interaction). Depending on the type of
crystal lattice there may exist different types of magnetic ground
state, often rather nontrivial, especially in frustrated lattices
containing e.g.\ magnetic triangles or tetrahedra as
building blocks.

\begin{figure}[ht]
  \centering
  \includegraphics[scale=1]{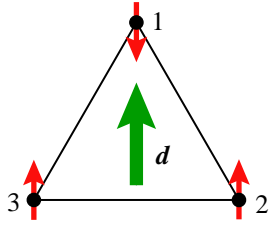}
  \caption{{\bf Electronic mechanism of dipole formation.} The formation
           of an electric dipole (green arrow) on a triangle
           of three spins (red arrow).}
  \label{fig:1}
\end{figure}

One can show~\cite{BBMK,BBMK2} that, depending on the magnetic
configuration, there can occur a spontaneous
charge redistribution in such a magnetic triangle, so that e.g.\ the electron density on site~1
belonging to the triangle (1,2,3) is
\begin{equation}
n_1 = 1 - 8 \left(\frac tU\right)^{\!3}\Bigl[\mbox{\boldmath{$S$}}_1\cdot(\mbox{\boldmath{$S$}}_2+\mbox{\boldmath{$S$}}_3) - 2\mbox{\boldmath{$S$}}_2\cdot\mbox{\boldmath{$S$}}_3\Bigr]\;
\label{eq:2}
\end{equation}
(in other spin textures there may appear spontaneous orbital currents \cite{BBMK,BBMK2} in such triangles.)
From this expression
one sees, in particular, that there should occur charge
redistribution for a triangle with two spins up and one down, Fig.~\ref{fig:1},
which would finally give a dipole moment
\begin{equation}
d \sim \mbox{\boldmath{$S$}}_1\cdot(\mbox{\boldmath{$S$}}_2+\mbox{\boldmath{$S$}}_3) - 2\mbox{\boldmath{$S$}}_2\cdot\mbox{\boldmath{$S$}}_3
\label{eq:3}
\end{equation}
shown in Fig.~\ref{fig:1} by a broad green arrow.

\begin{figure}[ht]
  \centering
  \includegraphics[scale=1]{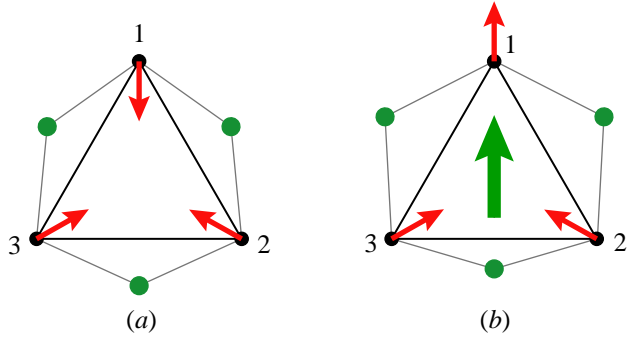}
  \caption{{\bf Magnetostriction mechanism of dipole formation.} Illustration
        of magnetostriction mechanism
        of the formation of an electric dipole (green arrow): the symmetric location
        of oxygens (green circles) for equivalent bonds~($a$)
        changes to an asymmetric one for spin configuration (red arrows) with
        different spin orientations on different bonds~($b$).}
  \label{fig:2}
\end{figure}

A similar expression describes also an electric dipole which can form
on a triangle due to the usual magnetostriction. One can illustrate
this e.g.\ on
the example of Fig.~\ref{fig:2}, see e.g.~\cite{mostovoy}, in which we show the
triangle (1,2,3) made by magnetic ions, with intermediate oxygens sitting
outside the triangle and forming a certain angle $M$--O--$M$.
For \hbox{3-in} spins, Fig.~\ref{fig:2}($a$), all three bonds
are equivalent, and all $M$--O--$M$ angles are the same.
However, in a configuration of Fig.~\ref{fig:2}($b$) (which,
according to Eq.~(\ref{eq:2}), would give a nonzero dipole
moment due to electronic mechanism), two bonds become ``more
ferromagnetic'', and the oxygens would shift as shown in Fig.~\ref{fig:2}($b$),
so as to
make the $M$--O--$M$ angle in the ``antiferromagnetic'' bond closer to 180
degrees, and in ``ferromagnetic'' bonds closer to 90 degrees;
according to the Goodenough--Kanamori--Anderson rules this would strengthen
the corresponding antiferromagnetic and ferromagnetic exchange and lead
to energy gain. As one sees from Fig.~\ref{fig:2}(b), such distortions
shift the centre of gravity of positive ($M$) and negative (O) charges and
thus would produce a dipole moment similar to that of Fig.~\ref{fig:1}. A similar
effect would also exist in a monopole configuration of spin ice, in
which on some bonds the spins are oriented ``ferromagnetic-like'' (e.g.\
on bonds with \hbox{2-in} spins), and on other bonds the spins are ``more antiferromagnetic''
(bonds with \hbox{1-in} and \hbox{1-out} spins).

\begin{figure}[ht]
  \centering
  \includegraphics[scale=1]{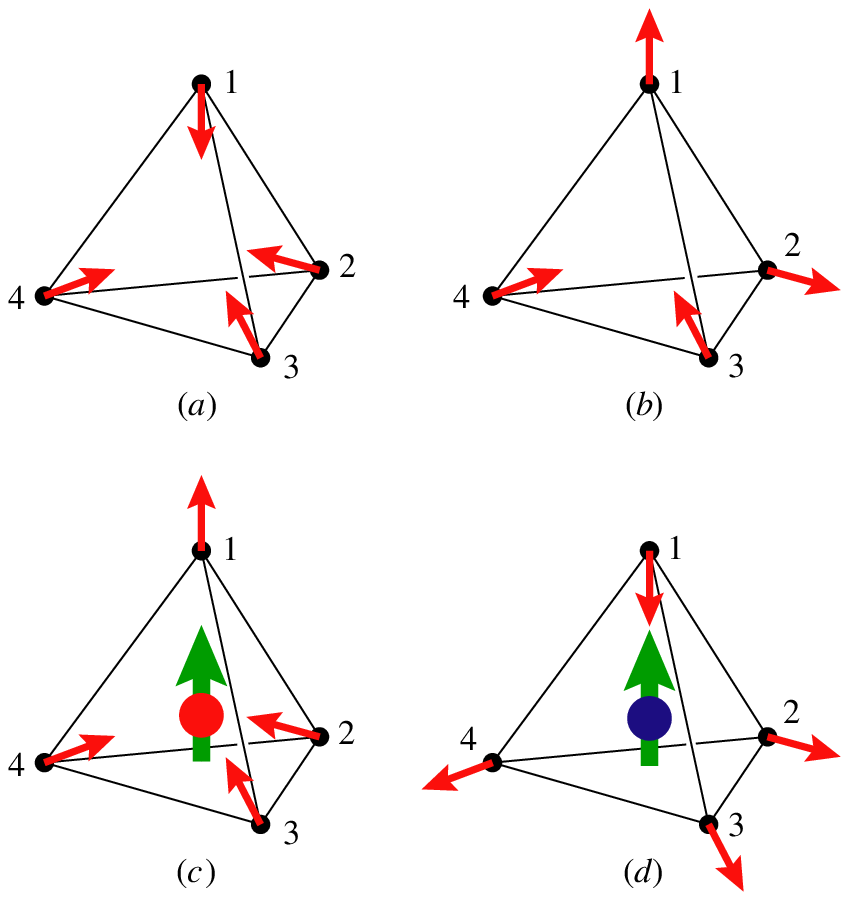}
  \caption{{\bf Formation of dipoles on monopoles.} Possible
     spin states (red arrows) in spin-ice-like systems,
     showing the formation of electric dipoles (broad green arrow)
     in monopole ($c$) and antimonopole ($d$) configurations
     (dipoles are absent in \hbox{4-in} ($a$) and \hbox{2-in/2-out} ($b$) states).
     Note that the direction of dipoles in cases ($c$), ($d$) is the
     same (in the direction of the ``special'' spin $\mbox{\boldmath{$S$}}_1$).}
  \label{fig:3}
\end{figure}

The expression (\ref{eq:3}) is the main expression, which gives the ``dipole on
monopole'' in spin ice. Indeed, when one considers three possible
configuration of a tetrahedron in spin ice, Fig.~\ref{fig:3}($a$) (\hbox{4-in} or
\hbox{4-out} state), the monopole configuration of Fig.~\ref{fig:3}($c,d$) (\hbox{3-in/1-out} or
\hbox{1-in/3-out}), and the basic spin ice configurations \hbox{2-in/2-out},
Fig.~\ref{fig:3}($b$), then, applying the expressions (\ref{eq:2}),~(\ref{eq:3}) to every triangle
constituting a tetrahedron, one can easily see that there would be no
net dipole moments in the cases of Fig.~\ref{fig:3}($a$) (\hbox{4-in} or \hbox{4-out}) and Fig.~\ref{fig:3}($b$)
(\hbox{2-in/2-out}), but there will appear a finite dipole moment in the case of
Fig.~\ref{fig:3}($c,d$), i.e.\ {\it there will appear an electric dipole on each magnetic
monopole in spin ice}.

The easiest way to check this is to start from the case \ref{fig:3}($a$), with
\hbox{4-in} spins. The total charge transfer e.g.\ on site~1 is
\begin{equation}
\delta n_1 \sim 2\mbox{\boldmath{$S$}}_1\cdot(\mbox{\boldmath{$S$}}_2 + \mbox{\boldmath{$S$}}_3 + \mbox{\boldmath{$S$}}_4) - 2(\mbox{\boldmath{$S$}}_2\cdot\mbox{\boldmath{$S$}}_3 + \mbox{\boldmath{$S$}}_2\cdot\mbox{\boldmath{$S$}}_4 + \mbox{\boldmath{$S$}}_3\cdot\mbox{\boldmath{$S$}}_4)\;.
\label{eq:4}
\end{equation}
For the 4-in state all the scalar products ($\mbox{\boldmath{$S$}}_i\cdot\mbox{\boldmath{$S$}}_j$) are equal, i.e.\
the charge redistribution, and with it the net dipole moment of the
tetrahedron is zero. (One can also use the condition $\mbox{\boldmath{$S$}}_1+\mbox{\boldmath{$S$}}_2+\mbox{\boldmath{$S$}}_3+\mbox{\boldmath{$S$}}_4
= \mbox{\boldmath{$0$}}$, valid in this case, to prove this; the fact that the dipole moment
is zero also follows just from the symmetry.)

However when we reverse the direction of one spin, e.g.\ $\mbox{\boldmath{$S$}}_1\to-\mbox{\boldmath{$S$}}_1$,
creating a \hbox{3-in/1-out} monopole configuration of Fig.~\ref{fig:3}($c$), the
first term in Eq.~(\ref{eq:4}) changes sign, and the resulting charge transfer
from sites 2, 3 and~4 to site~1 would be non-zero --- and there
will appear a dipole moment on such a tetrahedron, directed from the
centre of the tetrahedron to the site with the ``special spin'', in this
case to site~1 --- the broad green arrow in Fig.~\ref{fig:3}($c$)
(or in the opposite direction, depending on the specific situation ---
the sign of the hopping $t$ in Eq.~(\ref{eq:2}), or the details
of the exchange striction).  This conclusion,
shown in Figs.~\ref{fig:3}$(c,d)$, is actually the main result of this paper.

As the expressions (\ref{eq:2})--(\ref{eq:4}) for the charge redistribution and for the
dipole moment are even functions of spins~$\mbox{\boldmath{$S$}}$, the reversal of all
spins will not change the results. Thus the magnitude {\it and the
direction} of the electric dipole is the same for both the monopole
(\hbox{3-in/1-out})
and antimonopole  (\hbox{1-in/3-out}) configurations, Fig.~\ref{fig:3}($c$) and
\ref{fig:3}($d$): in both cases the dipole points in the direction of the
``special'' spin.

Similar considerations show that when we change the direction of one
more spin, e.g.\ $\mbox{\boldmath{$S$}}_2 \, {\to} \, -\mbox{\boldmath{$S$}}_2$, creating the \hbox{2-in/2-out} configuration of
Fig.~\ref{fig:3}($b$), various terms in Eq.~(\ref{eq:4}) again cancel, and such spin
configurations do not produce electric dipole. Thus, electric dipoles
appear in spin ice only on monopoles and antimonopoles.

\subsection*{Some consequences}
The appearance of electric dipoles on monopoles in spin ice could
have many consequences, some of which we now discuss. The main effect
would be the coupling of such dipoles to the dc or ac electric field,
\begin{equation}
{\cal E} = -\mbox{\boldmath{$d$}}\cdot\mbox{\boldmath{$E$}}\;.
\label{eq:5}
\end{equation}
This would give an electric activity to monopoles, would allow one to
influence them by external electric field, and would thus open a
new way to study and control such monopoles in spin ice. Due to this
coupling the monopoles would contribute to the dielectric function $\epsilon(\omega)$. Actually such
effect was observed in \cite{Saito}, where it was found that the
dielectric function has strong anomalies in Dy$_2$Ti$_2$O$_7$ in the magnetic
field in the [111] direction when the system approaches a transition to
the saturated state at $H \sim 1\,\rm T$~\cite{Aoki}. The mechanism of these
anomalies was not discussed in \cite{Saito}, but one can connect it
with the proliferation of monopoles and antimonopoles, with the corresponding
electric dipoles on each of them, in approaching this transition.

\begin{figure}[ht]
  \centering
  \includegraphics[scale=1]{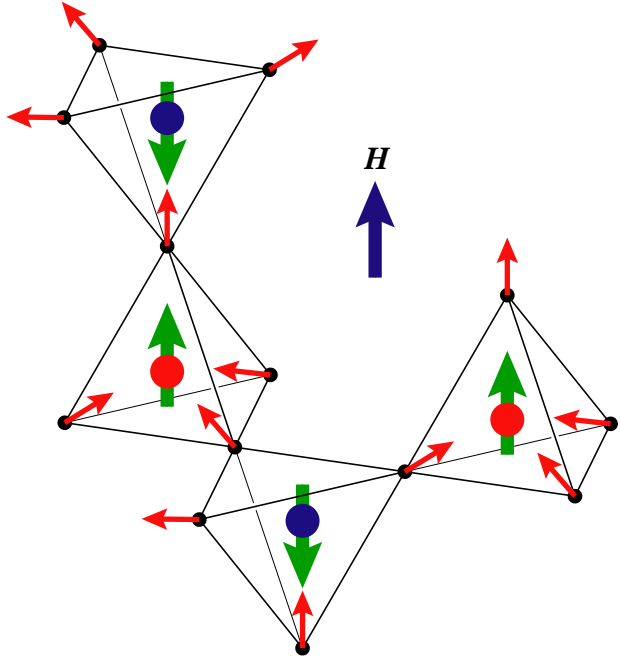}
  \caption{{\bf Ordered spin configuration in spin ice
        in a strong [111] magnetic field.} This structure can be seen as an ordered array
        of monopoles and antimonopoles; simultaneously it is antiferroelectric
        (electric dipoles are shown by broad green arrows).}
  \label{fig:4}
\end{figure}

The saturated state in this situation, shown in Fig.~\ref{fig:4}, has the form
of staggered monopoles--antimonopoles at every tetrahedron. From our
results presented above, we conclude that in this state there would also be
electric dipoles at every tetrahedron, shown in Fig.~\ref{fig:4} by thick
green arrows. We see thus that this saturated state in a strong enough
[111] magnetic field would simultaneously be antiferroelectric. Thus
one can also associate the anomalies observed in \cite{Saito} in
$\epsilon(\omega)$ in approaching this state as the anomalies at the
antiferroelectric transition.

Yet another consequence of the appearance of dipoles on monopoles
could be the possibility of changing the activation energy for creating
such monopoles by electric field: the excitation energy of a
monopole, or the monopole--antimonopole pair would be
\begin{equation}
\Delta = \Delta_0 - \mbox{\boldmath{$d$}}\cdot\mbox{\boldmath{$E$}}\;.
\end{equation}
Correspondingly, depending on the relative orientation of $\mbox{\boldmath{$d$}}$ and $\mbox{\boldmath{$E$}}$,
the excitation energy can both increase and decrease, but one can
always find configurations of monopoles for which the energy
would decrease. One should then be able to see this change of
activation energy in thermodynamic and magnetic properties, such as
specific heat, etc.

\begin{figure}[ht]
  \centering
  \includegraphics[scale=1]{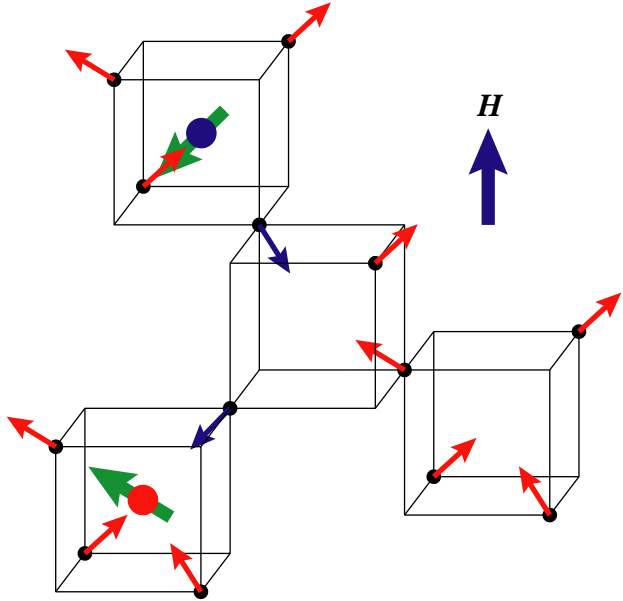}
  \caption{{\bf Possible monopole--antimonopole
        pair in strong [001] magnetic field.}
        The $z$-component of electric dipoles (broad green arrows) on
        monopoles is pointing up, and on antimonopoles down.
        The perpendicular components of $\mbox{\boldmath{$d$}}$ point
        in random [110] and [$1\bar 10$] directions.
        Blue arrows show spins inverted in creating
        and moving apart monopole and antimonopole.}
  \label{fig:5}
\end{figure}

The orientation of electric dipoles depends on the particular situation.
One can easily see that in the absence of magnetic fields, for
completely ``free'', random spin ice, in general the orientation of dipoles
on monopole excitations is
random, in all [111] directions. But, for example, in strong
enough [001] magnetic field, in which the spin ice state is
ordered, Fig.~\ref{fig:5}, the monopoles and antimonopoles would have the
$z$-components of dipoles respectively positive and negative, $d^z
\hbox{(monopoles)}>\nobreak0$, $d^z\hbox{(antimonopoles)} <\nobreak 0$, while the perpendicular
projections of $\mbox{\boldmath{$d$}}$ would be random.
Similarly, in the [110] field \cite{fennell} the $xy$-projection of
dipole moments will be parallel to the field, $d_{xy} \parallel [110]$.

\begin{figure}[ht]
  \centering
  \includegraphics[scale=1]{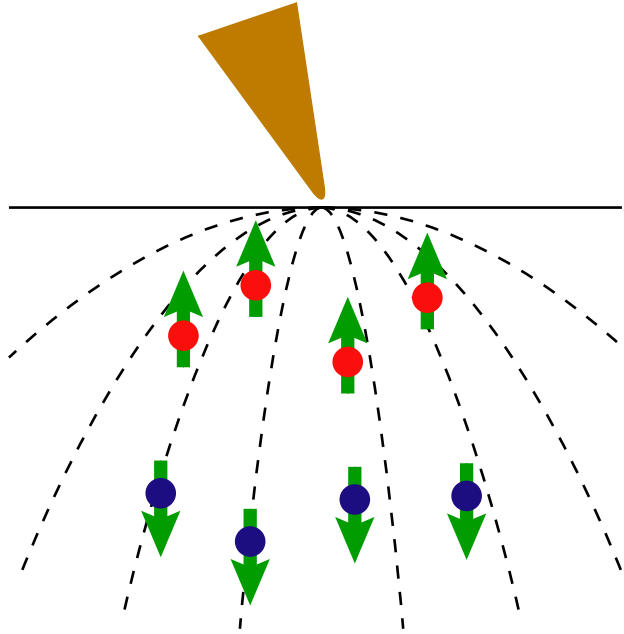}
  \caption{{\bf Separation of monopoles and antimonopoles.} The behaviour of monopoles and antimonopoles
        with respective electric dipoles (green arrows) in spin ice in [111] or [001]
        magnetic field in the inhomogeneous electric field (dashed lines)
        created by a tip (brown) with electric voltage.}
  \label{fig:6}
\end{figure}

Yet another effect could appear in an inhomogeneous electric field,
created for example close to a tip with electric voltage applied to it, in
an experimental set-up shown in Fig.~\ref{fig:6} (cf.\ e.g.\ the study of N\'eel
domain walls in a ferromagnet, which also develop electric
polarization and which can be influenced by inhomogeneous electric field
\cite{mgu}). As always, the electric dipoles would move in $\hbox{grad}\,\mbox{\boldmath{$E$}}$,
with positive dipoles e.g.\ being attracted to the region of stronger field
and negative ones repelled from it. One can use this effect to
``separate'' monopoles from antimonopoles. Thus, as is clear from
Fig.~\ref{fig:4}, in a [111] magnetic field, e.g.\ in the phase of ``kagome ice''
\cite{Aoki}, the ``favourable'' monopoles would have dipole moments
up, and antimonopoles down, so that the monopoles would be
attracted to the tip, to the region of stronger electric field, and
antimonopoles would be repelled from the tip. Similarly, the
monopole--antimonopole separation could be reached in a [001] magnetic
field, in which, as we have argued above, Fig.~\ref{fig:5}, monopoles have $d^z>0$, and
antimonopoles have~$d^z<0$.

The magnitude of the dipoles created on monopoles, and the corresponding
strengths of their interaction with electric field, depend on the
detailed mechanism of their creation and on the specific properties
of a given material.
One should think that in real
spin ice materials, in which the hopping of $f$-electrons is rather
small, it is the magnetostriction mechanism of the dipole formation on monopoles
and antimonopoles that would be the dominant~one. In this case
one could make a
crude estimate based on the interaction~(\ref{eq:5}).  If the shifts of ions
$u$ due to striction would be e.g.\ of order $0.01\,$\AA,
then the change of the energy ${\cal E} = -\mbox{\boldmath{$d$}}\cdot\mbox{\boldmath{$E$}} = -euE$ in a
field $E \sim 10^{5}\,\rm V/cm$ would be $\sim0.1\,\rm K$ --- which would lead to
measurable effects, as the typical excitation energy of monopoles
in spin ice is $\sim1\,\rm K$ \cite{castelnovo,bramwell}.
We would get effects of the same order of magnitude for the distortions $u\sim10^{-3}\,$\AA\
in a field~$\sim10^6\,\rm V/cm$.

\section*{Discussion}
Summarizing, we demonstrated that there should appear real electric
dipoles on magnetic monopoles in spin ice.  Creation of such
dipoles may lead to many experimental consequences, some of which
were discussed above. They can open new ways to study and to
manipulate these exciting new objects --- magnetic monopoles in a
solid. We also see that the close connection between electric and
magnetic phenomena, which lies at the hart of modern physics,
extends in this case even further than one thought: in these systems one can have not only
magnetic charges instead of electric ones,
and ``magnetricity'' instead of electricity, but, similar to electrons
which have {\it electric charge} and {\it magnetic dipole} (spin), magnetic
monopoles in spin ice will have {\it magnetic charge} and {\it electric dipole}.

\section*{Acknowledgements}
I am grateful to C.~D.~Batista, S.-W.~Cheong and M.~J.~P.~Gingras for useful discussions.
This work was supported by the German project SFB 608 and by the
European project SOPRANO.

%
%
%

\end{document}